\documentclass[aps,pra,twocolumn,superscriptaddress]{revtex4-2}

\usepackage{graphicx}
\usepackage{siunitx}
\usepackage[hyperfootnotes=false,linkbordercolor={0 .298 .6}, citebordercolor={.8 0 0}, urlbordercolor={0 .298 .6}]{hyperref}

\hypersetup{
	colorlinks,
	linkcolor=[rgb]{0,.298,.6},
	citecolor=[rgb]{0,.298,.6},
	urlcolor=[rgb]{0,.298,.6}
}

\newcommand{\invcm}{{cm^{-1}}}
\newcommand{\figFolder}{.} 

\begin{document}

\title{Two-photon excitation and absorption spectroscopy of gaseous and supercritical xenon}

\author{Thilo vom H{\"o}vel}\email[]{hoevel@iap.uni-bonn.de}

\author{Franz Huybrechts}
\author{Eric Boltersdorf}
\author{Christian Wahl}
\author{Frank Vewinger}
\author{Martin Weitz}
\affiliation{Institut f{\"u}r Angewandte Physik, Universit{\"a}t Bonn, Wegelerstra{\ss}e 8, 53115 Bonn, Germany}

\date{\today}

\begin{abstract}
	
	Spectroscopy of gases under high-pressure conditions is of interest in various fields such as plasma physics and astrophysics. Recently, it has also been proposed to utilize a high-pressure noble gas environment as a thermalization medium to extend the wavelength range of photon Bose-Einstein condensates to the vacuum-ultraviolet regime, from the presently accessible visible and near-infrared spectral regimes. In this work, we report on experimental results of two-photon spectroscopy of gaseous and supercritical xenon for pressures as high as \SI{95}{\bar}, probing the transitions from the $5p^6$ electronic ground-state to the $5p^56p$ and $5p^56p^\prime$ excited-state configurations. Aiming at the exploration of possible pumping schemes for future vacuum-ultraviolet photon condensates, we have recorded degenerate two-photon excitation spectra of such dense xenon samples. In further measurements, we have investigated whether irradiation of an auxiliary light field can enhance the reabsorption of the emission on the second excimer continuum of xenon, which is subject to a large Stokes shift. To this end, absorption measurements have been conducted, driving the $5p^6 \rightarrow 5p^56p$ two-photon transitions nondegenerately.	
 
\end{abstract}

\maketitle

\section{Introduction}

	The spectra of gases under high-pressure conditions are known to show a range of universal features, such as the applicability of the thermodynamic Kennard-Stepanov relation and Kasha's rule \cite{KennardFluorescence1918, StepanovFluorescence1957, SawickiKennardStepanov1996, KashaCharacterization1950, MoroshkinKennardStepanov2014}, caused by a thermalization of internal energy levels induced by frequent collisions between constituent particles. Accordingly, it seems natural to exploit these universal features for the thermalization of light \cite{KlaersThermalization2010}. In a thermally equilibrated gas that is sufficiently cold and dense for the thermal de Broglie wave packets to spatially overlap, quantum statistical effects emerge. For integer-spin particles, Bose-Einstein condensation to the lowest-energy eigenstate can then occur, as has been observed among others for ultracold atomic gases \cite{AndersonBEC1995, DavisBEC1995}, exciton polaritons \cite{KasprzakExcitonPolariton2006, SuExcitonPolariton2020}, and photons confined to optical microcavities \cite{KlaersCondensation2010, MarelicCondensation2015, GrevelingDensity2018} or fibers \cite{WeillFibers2019}. 
	
	The centerpiece of a typical experimental platform for photon Bose-Einstein condensation is a dye solution-filled optical microcavity with a very short (wavelength-size) mirror spacing, introducing a low-frequency cutoff for cavity photons, as well as a matterlike (i.e., quadratic) dispersion relation. Photons in the visible spectral range near \SI{580}{\nm} wavelength thermalize by repeated absorption-reemission processes to the (rovibrational) temperature of the dye molecules, which are at room temperature. The dye molecules fulfill the Kennard-Stepanov relation, a Boltzmann-like thermodynamic frequency scaling between the absorption and the emission spectral profiles. This universal relation applies to systems where the rovibrational manifolds of both the lower and the upper electronic state are in thermal equilibrium \cite{BandMcCumber1988, MartinMcCumber2006, MoroshkinKennardStepanov2014, OckenfelsSpectroscopy2022}, as is the case in this system due to frequent collisions of dye molecules with surrounding solvent molecules occurring on a \SI{e-14}{\second} timescale, much shorter than the radiative lifetime of the upper electronic state. In contrast to a usual laser, no population inversion of the active medium is required to generate coherent optical emission \cite{SchmittCoherence2016} and spontaneous emission can be retrapped in the microcavity.
	
	Bose-Einstein condensates of photons are accordingly expected to be an attractive platform for a coherent light source in the vacuum-ultraviolet spectral regime (VUV), where the operation of a laser requires high pump intensities, given the $1/\omega^3$ scaling of upper electronic state lifetimes, with $\omega$ as the optical frequency. Since in this spectral regime dye molecules cannot be employed as the thermalization medium due to a lack of suitable transitions, earlier work \cite{WahlAbsorption2018, WahlXenon2021} suggested to use noble gases exhibiting closed electronic transitions in this wavelength range, e.g., around \SI{147}{\nm} wavelength for the xenon $5p^6 \rightarrow 5p^56s$ transition at high pressures. The frequent interatomic collisions occurring in such dense gas samples can lead to the thermalization of the quasimolecular collisional manifolds in both ground and electronically excited states, expected to induce a Kennard-Stepanov frequency scaling between absorption and emission spectral profiles \cite{MoroshkinKennardStepanov2014}, as necessary to enable the thermalization of photons. Other than in dilute gases, at pressures in the \SI{100}{\bar} regime, the collisional line broadening of gas atoms typically becomes comparable to the thermal energy in frequency units $k_{\text{B}}T/\hbar$, and thermalization of photons can become effective.

\begin{figure}
		\includegraphics[]{\figFolder/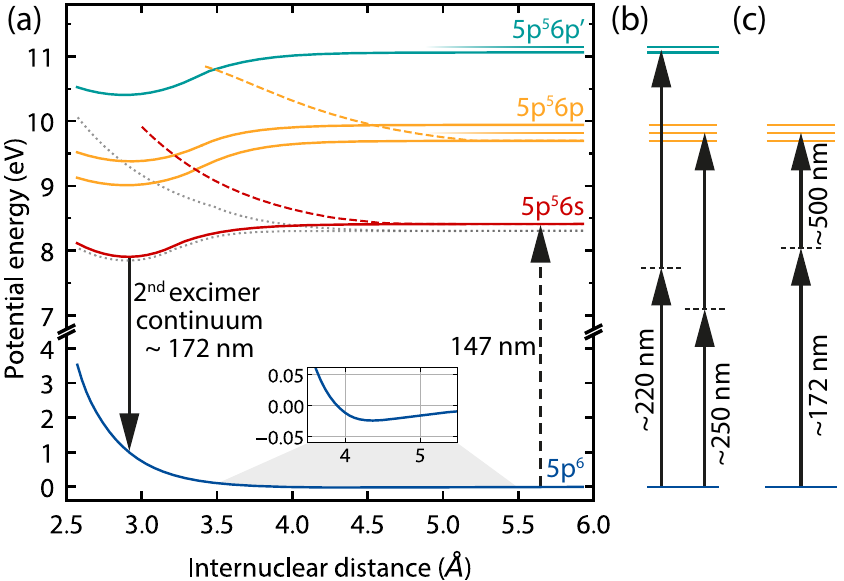}
		\caption{(a) Schematic of the potential-energy curves in the xenon dimer system. For the sake of clarity, only relevant states are shown; for a complete visualization, see \cite{BoweringCollisionalState1986, SelgPotentials2003, NISTenergyLevels2023}. The quasimolecular ground state, constituted by two colliding xenon atoms in the $5p^6$ state, is mostly repulsive (the inset shows a shallow well around \SI{4.3}{\angstrom}). The first excited state, constituted by one atom each in the $5p^6$ and $5p^56s$ states, exhibits a pronounced potential well around \SI{2.9}{\angstrom}, leading to the formation of bound excimers. For those quasimolecular states of the higher-lying $5p^56p$ and $5p^56p'$ configurations whose exact potential-energy curve shape is not covered by the literature, horizontal lines indicate the asymptotic energies. The dotted lines refer to potential energy curves associated with the metastable $(J = 2)$ component of the $5p^56s$ state. [(b), (c)] Electronic states within the $5p^56p$ and $5p^56p'$ configurations can be excited by two-photon transitions in a (b) degenerate or (c) nondegenerate manner.
			\label{fig: potentialCurves}}
	\end{figure}

	A further demand on a thermalization medium is a large reabsorption probability for radiation in the spectral emission region, such that in future microcavity-based experiments the confined radiation is absorbed and reemitted multiple times, as necessary to establish thermal contact between the photons and the medium. For xenon, a schematic of the quasimolecular potential curves of the relevant electronic levels in a binary collision picture is shown in Fig. \ref{fig: potentialCurves}(a). The lowest-energy absorption line ($5p^6 \rightarrow 5p^56s$) is near \SI{147}{\nm} wavelength \cite{SelgPotentials2003}. For gas pressures exceeding a few-hundred millibar, emission is observed on the so-called second excimer continuum at around \SI{172}{\nm} wavelength, subject to a Stokes shift \cite{StokesRefrangibility1852} of about \SI{25}{\nm}, as understood from the quasimolecular potential curves. This large energetic gap between absorption and emission leads to a small reabsorption probability for radiation emitted on the second excimer continuum. A closing of this energetic gap could be enabled by an auxiliary light field, facilitating two-photon transitions from the electronic ground state to the higher-energetic states of the $5p^56p$ or $5p^56p^\prime$ configurations with even total angular momentum [Figs. \ref{fig: potentialCurves}(b) and \ref{fig: potentialCurves}(c)]. On a fine-structure level, the $5p^56s$ electronic state consists of two components: one with $J = 1$, linked by an optically allowed transition to the $5p^6$ $(J = 0)$ ground state, and a metastable one with $J = 2$, which in the asymptotic limit is the lowest-energetic atomic fine-structure component. While direct laser excitation of the metastable  $5p^56s$ $(J = 2)$ state from the ground state is dipole forbidden, the metastable level can be populated via potential curve crossings of excited quasimolecular states in dense gas systems with frequent interconstituent collisions.
		
	Available spectroscopic data of samples around the regime in between usual gas and liquid phase conditions, where the collisional spectral line broadening is of the order of the thermal energy in frequency units, are relatively scarce. In earlier work, absorption and emission VUV spectra of xenon have been recorded at pressures of up to \SI{130}{\bar} \cite{BorovichAbsorption1973, WahlAbsorption2018, WahlXenon2021}. Two-photon excitation spectra of the $5p^6 \rightarrow 5p^56p$ and $5p^56p^\prime$ transitions of xenon have been recorded in a pressure range of up to \SI{13}{\bar} \cite{GornikTwoPhoton1981, RaymondTwoPhoton1984, BoweringCollisionalState1986, BoweringCollisionalLifetimes1986, WhiteheadDeactivation1995}, with the excitation usually detected by monitoring the rates of the second excimer emission near \SI{172}{\nm} wavelength or those of the intermediate infrared emission. For a fixed excitation wavelength of \SI{266}{\nm}, line profiles of the second excimer emission have been analyzed at xenon pressures as high as \SI{180}{\bar} \cite{WahlXenon2021}. In the moderate-pressure regime, multiphoton ionization spectroscopy has been performed for both atomic and diatomic xenon \cite{AronMultiphoton1977}.
	
	The aim of the present work is to obtain (V)UV spectroscopic data of such dense xenon ensembles. We report on the observation of excitation spectra of degenerate $5p^6 \rightarrow 5p^56p$ and $5p^56p^\prime$ two-photon transitions of xenon at pressures of up to \SI{95}{\bar}. While, in the gaseous regime, the spectral components of the transitions to the individual upper states of the $5p^56p$ and $5p^56p^\prime$ configurations remain spectrally resolved, at higher pressures upon which the supercritical regime is reached, the observed pressure broadening exceeds the splitting between the individual configurations. Moreover, we present $5p^6 \rightarrow 5p^56p$ two-photon absorption spectra of xenon, driven nondegenerately \cite{SadeghNonDegenerate2019} with radiation near the second excimer continuum around \SI{172}{\nm} and visible spectral range photons near \SI{500}{\nm} wavelength while monitoring the absorption of the light field around \SI{172}{\nm}. Our results suggest that the large Stokes shift between absorption and emission of the xenon $5p^6 \rightarrow 5p^56s$ single-photon transition can be compensated by irradiation with an auxiliary light field, enabling the reabsorption of radiation emitted on the excimer transition as required for photon thermalization.
	
	In the following, Sec. \ref{sec: experimentalEnvironment} describes the experimental setup used here and Sec. \ref{sec: ExcitationSpectroscopy} presents two-photon excitation spectroscopy studies of dense xenon samples. Subsequently, measurements exploring the feasibility of Stokes shift compensation by irradiation of an auxiliary light field are reported in Sec. \ref{sec: AbsorptionSpectroscopy}. Finally, Sec. \ref{sec: conclusions} gives conclusions and an outlook.

\section{Experimental Environment}
\label{sec: experimentalEnvironment}

	\begin{figure*}
		\includegraphics[]{\figFolder/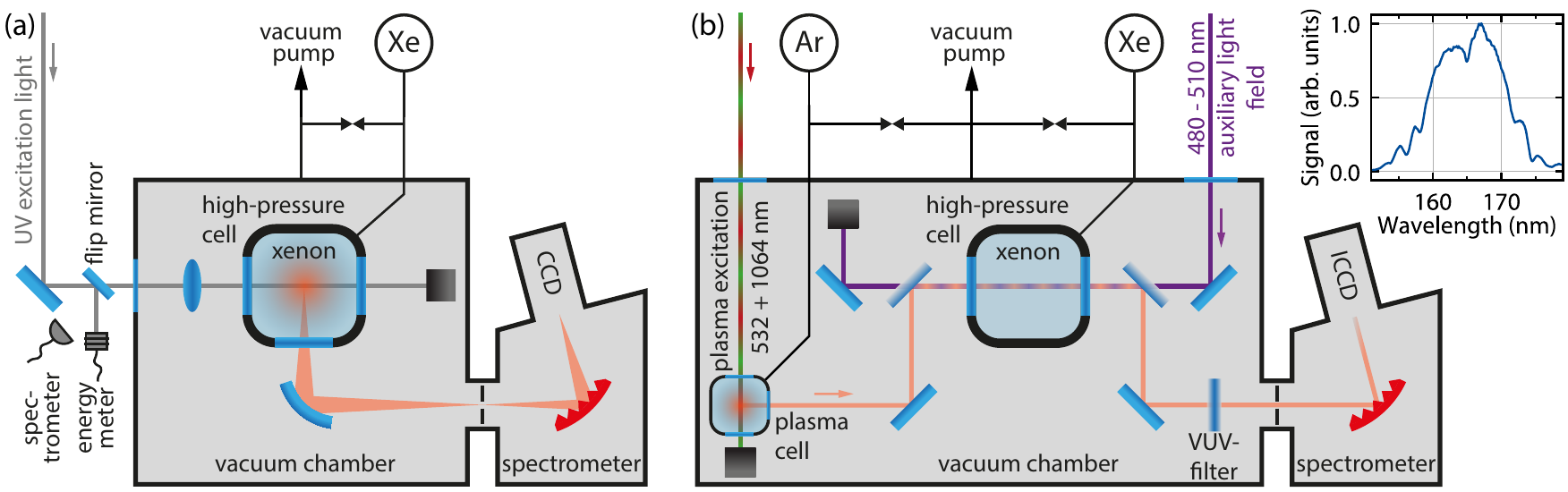}%
		\caption{(a) Setup for two-photon excitation spectroscopy. The exciting UV beam is focused into the center of a xenon-filled high-pressure cell and the resulting fluorescent radiation in the vacuum-ultraviolet spectral regime is collected with a concave mirror and guided to a grating spectrometer. (b) Setup for two-photon absorption spectroscopy. Broadband VUV radiation from a laser-induced plasma is overlapped with a counterpropagating beam of tunable laser radiation near \SI{500}{\nm} wavelength and guided through a xenon-filled high-pressure cell. The transmitted VUV radiation is analyzed with a spectrometer comprising an ICCD camera, allowing for time-resolved detection. The inset shows a spectrum of the VUV beam generated in the laser-induced plasma as detected by the ICCD-based spectrometer in the absence of xenon in the high-pressure cell.
			\label{fig: experimentalEnvironment}}
	\end{figure*}

	Our experimental instrumentation is based on a setup described in earlier work \cite{WahlAbsorption2018, WahlXenon2021}. The centerpiece of the setup is high-pressure gas cells made from stainless steel, equipped with MgF$_2$ windows for optical access in the VUV spectral regime and an optical path length of \SI{2}{\cm}. For all measurements described in this work, we use xenon (grade 4.8, Air Liquide). The high-pressure cell is placed in a vacuum chamber, along with all optics required to guide VUV radiation to and from the sample. The vacuum chamber is evacuated to below \SI{5e-5}{\milli\bar} to suppress absorption of short-wavelength radiation by atmospheric gases \cite{WatanabeOxygen1953, WatanabeWaterVapor1953, ZaidelVUV1970}. 
		 
	To generate tunable UV radiation, we use a Q-switched pulsed laser (Spectra Physics, model {Quanta Ray PRO 250-10}), providing radiation near \SI{1064}{\nm} wavelength with a pulse energy of \SI{1400}{mJ}, a pulse length of \SI{10}{\ns}, and a repetition rate of \SI{10}{\hertz}. After second-harmonic and subsequent sum-frequency generation, the resulting light near \SI{355}{\nm} wavelength is used to pump an optical parametric oscillator (OPO; GWU, model {primoScan ULD400}). For the excitation spectroscopy measurements, its tunable emission in the visible spectral range is frequency doubled and guided into the vacuum chamber through a viewport [Fig. \ref{fig: experimentalEnvironment}(a)]. To monitor the excitation beam wavelength, radiation scattered off the viewport is guided to a fiber-coupled spectrometer (Thorlabs, model {CCS200}). Inside the vacuum chamber, the excitation beam is focused into the center of the high-pressure cell by a lens with \SI{25}{\centi\meter} focal length to a beam diameter near \SI{10}{\micro\meter}. For the detection of atoms transferred to the $5p^56p$ and $5p^56p^\prime$ excited states by two-photon transitions [see Fig. \ref{fig: potentialCurves}(b)], part of the ensuing secondary emission near \SI{172}{\nm} wavelength leaving the high-pressure cell through a window in a direction orthogonal to the excitation beam is collected with a concave mirror and guided onto the entrance slit of a commercial VUV spectrometer (McPherson, model {234/302}). This consists of a concave grating and an open-nose CCD camera (Andor, model {iDus 420}), whose sensor chip can be cooled down to \SI{-70}{\degreeCelsius}. For the chosen width of the entrance slit of \SI{10}{\um}, the spectral resolution of the spectrometer is about \SI{0.1}{\nm}.
	
	Figure \ref{fig: experimentalEnvironment}(b) shows the setup used for the absorption spectroscopy measurements. Here, a laser-induced plasma (LIP \cite{LaportePlasma1987, PalmaPlasma2007, KakuPlasma2012}) is used as a broadband light source in the wavelength range near the second excimer continuum, implemented by focusing a part of the emission of the pump laser and its second harmonic (at 1064 and \SI{532}{\nm} wavelength, respectively) into a stainless-steel cell filled with argon gas at a pressure of about \SI{1}{\bar}. After the radiation generated in the LIP is collected and collimated, it is guided through a high-pressure spectroscopy cell of the same type as used in the above-described excitation spectroscopy setup via two dichroic mirrors (each exhibiting a reflectivity maximum centered around \SI{165}{\nm} wavelength with a \SI{20}{\nm} spectral width). The transmitted VUV radiation then passes a VUV bandpass filter (maximum transmissivity at \SI{150}{\nm}, spectral width about \SI{50}{\nm}) and is focused onto the entrance slit of a commercial VUV spectrometer (H+P spectroscopy, model {easyLight}). The latter comprises a concave grating and an intensified CCD camera (Andor, model {iStar 320T}), allowing for gated detection with gate widths as short as \SI{2}{\nano\second}, much shorter than the length of the plasma pulse, which is in the order of a few-hundred nanoseconds. To allow for a driving of the $5p^6 \rightarrow 5p^56p$ two-photon transitions via VUV radiation near \SI{172}{\nm} on its passage through the xenon-filled spectroscopy cell [see Fig. \ref{fig: potentialCurves}(c)], the VUV beam is spatially overlapped with an additional counterpropagating beam, exploiting the dichroic mirrors' transmissivity for visible light. This beam is tuned to near \SI{500}{\nm} wavelength, well within the tuning range of our OPO source. A spectrum of the VUV beam generated in the LIP, as recorded with the intensified CCD camera, is shown in the inset of Fig. \ref{fig: experimentalEnvironment}(b). The observed spectrum is significantly affected by the reflection and transmission profiles of the dichroic mirrors and the VUV filter, both of which contribute to the required high suppression of the residual visible and infrared scattered radiation. The width of the obtained spectrum is sufficient to cover the wavelength range in which VUV absorption is expected.
	
\section{Two-Photon Excitation Spectroscopy Measurements}
\label{sec: ExcitationSpectroscopy}

\begin{figure}
		\includegraphics[]{\figFolder/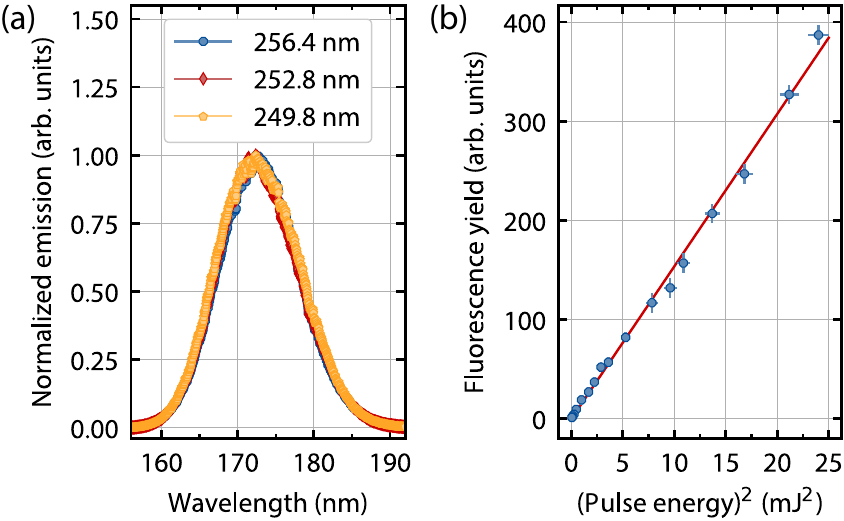}%
		\caption{(a) Xenon emission spectra recorded at a pressure of \SI{20}{\bar} for different excitation wavelengths driving the $5p^6 \rightarrow 5p^56p$ two-photon transitions. (b) Relation between the fluorescence yield (data points) and the square of the pulse energy of the excitation laser beam, along with a linear fit (line), for a fixed excitation wavelength of \SI{256.5}{\nm}.
			\label{fig: redistributionScalingArrangement}}
	\end{figure}

	We have studied degenerate two-photon excitation of the xenon $5p^6 \rightarrow 5p^56p$ transition, driven by two photons of the same wavelength [see Fig. \ref{fig: potentialCurves}(b)]. Subsequent deexcitation to the intermediate $5p^56s$ configuration, or the associated strongly bound molecular dimer levels of this excimer system, is possible either radiatively by emission of an infrared photon or by collisional deactivation, via the indicated crossings of the quasimolecular levels in Fig. \ref{fig: potentialCurves}(a). Reported collisional deactivation rates are \SIrange{6}{120}{\times10^{-12}\cubic\cm\per\second} \cite{BruceRates1990}, depending on the particular level of the $5p^56p$ configuration, which at room temperature corresponds to a rate of \SIrange{0.14}{2.8}{\per\bar\per\nano\second}. In the xenon pressure range of 10 to \SI{80}{\bar} covered by the present work, this causes the quench rate to clearly exceed that of spontaneous emission (around \SI{30}{\ns} natural lifetime) and hence we expect the transfer to the $5p^56s$ configuration to be dominated by collisional deexcitation. From this state, deexcitation to the $5p^6$ ground state occurs radiatively, by emission of photons on the second excimer continuum around \SI{172}{\nm} wavelength \cite{SieckEmission1968, BrodmannEmission1978, GornikTwoPhoton1981}. Figure \ref{fig: redistributionScalingArrangement}(a) shows experimental spectra of the laser-induced fluorescence for the three different excitation wavelengths \SI{249.8}{}, \SI{252.8}{}, and \SI{256.4}{\nm}, corresponding to a tuning near the $5p^6 \rightarrow 5p^56p[1/2]_0$, $5p^56p[3/2]_2$, and $5p^56p[5/2]_2$ transitions, respectively, recorded at a xenon pressure of \SI{20}{\bar}. Within the experimental accuracy, the observed emission spectra well agree in all three cases and do not or only weakly depend on the excitation mechanism, as expected from Kasha's rule, and can be identified as the familiar emission spectra of the second excimer continuum. Next, Fig. \ref{fig: redistributionScalingArrangement}(b) shows the variation of the spectrally integrated fluorescence yield on the square of the used pulse energy for an excitation wavelength of \SI{256.5}{\nm}. The data on this scale can be well fitted with a linear function, as expected for a two-photon excitation \cite{GoeppertMayerQuadratic1931, BloembergenQuadratic1976}.
	
\begin{figure}
		\includegraphics[]{\figFolder/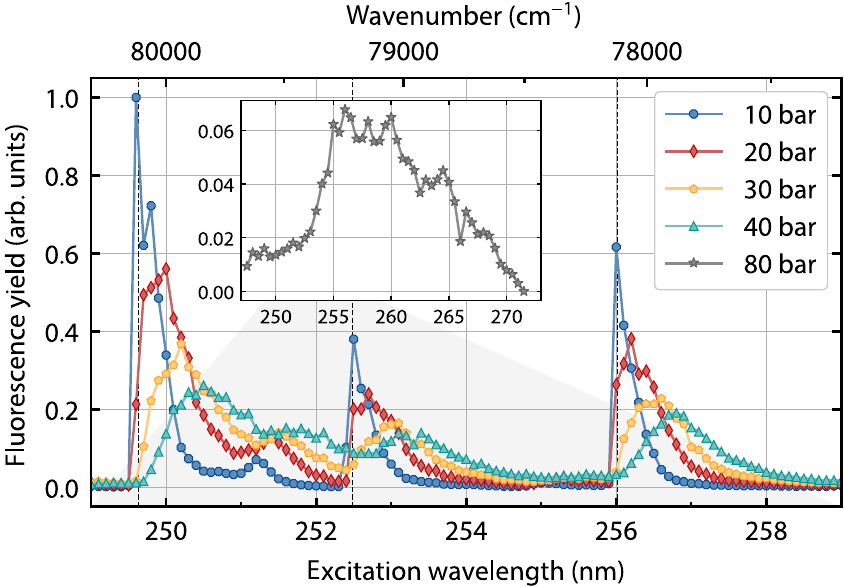}%
		\caption{Excitation spectra of the $5p^6 \rightarrow 5p^56p$ two-photon transitions in xenon. The main plot shows the relation between the spectrally integrated fluorescence yield and the wavelength of the exciting laser beam for different xenon pressures in the gaseous phase. The vertical dashed lines indicate the expected spectral positions of the $5p^6 \rightarrow 5p^56p[1/2]_0$, $5p^56p[3/2]_2$, and $5p^56p[5/2]_2$ transitions (from left to right) of the atomic xenon system \cite{NISTenergyLevels2023}. The inset shows a data set recorded at \SI{80}{\bar} pressure, well within the supercritical regime. Note the extended wavelength range and different vertical scale of the inset as compared to the main plot.
			\label{fig: 6p_resonances}}
	\end{figure}

	In the following, excitation spectroscopy measurements are presented, aimed at determining the fluorescence yield. The measurements have been conducted at a fixed pulse energy of \SI{2}{\milli\joule}, corresponding to a \SI{2.5}{\kilo\joule\per\cm\squared} energy density in the focus. Figure \ref{fig: 6p_resonances} shows the relation between the spectrally integrated fluorescence yield and the wavelength of the exciting laser beam for different xenon pressures. The main plot shows data for pressures in the range from 10 to \SI{40}{\bar}, i.e., with the sample in the gaseous phase, and the vertical dashed lines illustrate the expected positions for the undisturbed $5p^6 \rightarrow 5p^56p[1/2]_0$, $5p^56p[3/2]_2$, and $5p^56p[5/2]_2$ atomic transitions at \SI{80119.0}{}, \SI{79212.5}{}, and \SI{78119.8}{\per\centi\metre} (\SI{249.6}, {252.5}, and \SI{256.0}{\nm}), respectively; see the wave-number scale at the top of the figure \cite{NISTenergyLevels2023}. In the data set recorded at the lowest pressure of \SI{10}{\bar}, the corresponding transitions are clearly resolved experimentally, with three distinct peaks at around \SI{249.6}{}, \SI{252.5}{}, and \SI{256.0}{\nm} wavelength. In this regime, one observes a very pronounced asymmetry of the three resonances. The blue edges are very steep and their spectral positions coincide with those of the undisturbed atomic transitions to within our \SI{0.1}{\nm} experimental wavelength accuracy. In contrast, on their red edges, the observed features exhibit a broadened shape. Strongly asymmetric line shapes have also been observed in earlier work investigating the $5p^6 \rightarrow 5p^56p$ transition manifold at lower gas pressures and can be ascribed to downward-bending potential-energy curves of the $5p^56p \, (J = 0, 2)$ states \cite{GornikTwoPhoton1981}.
	
	Around \SI{251.2}{\nm}, an additional feature can be observed, its spectral position coinciding quite well with the $5p^6 \rightarrow 5p^55d[1/2]_0$ transition expected at \SI{79771.3}{\per\centi\metre} (\SI{250.7}{\nm}), which for atoms is dipole forbidden, but has been observed earlier and attributed to a pressure-induced transition \cite{BoeweringPressure1983}.

	With increasing pressure, we observe both a broadening and a shift of the resonances. The observed pressure shift of the $5p^6 \rightarrow 5p^56p[1/2]_0$, $5p^56p[3/2]_2$, and $5p^56p[5/2]_2$ resonances is \SI{9.3 \pm 1.4}{}, \SI{8.3 \pm 1.2}{}, and \SI{8.5 \pm 1.1}{\invcm/\bar} toward lower wave numbers, respectively. At the highest pressure in the gaseous phase investigated here of \SI{40}{\bar}, the peaks partly overlap.
	
	The inset in Fig. \ref{fig: 6p_resonances} shows a two-photon excitation spectrum recorded at a pressure of \SI{80}{\bar} (note the different wavelength scale), well within the supercritical phase, the transition to which occurs at a pressure of \SI{58.4}{\bar} at room temperature \cite{NISTchemistryWeb2023}. At this point, the extent of the pressure broadening renders resolving the individual spectral components of the two-photon transitions to the $5p^56p$ configuration unfeasible. The distinct change of spectral features upon crossing into the supercritical regime has been observed earlier \cite{WahlXenon2021}.
	
\begin{figure}
		\includegraphics[]{\figFolder/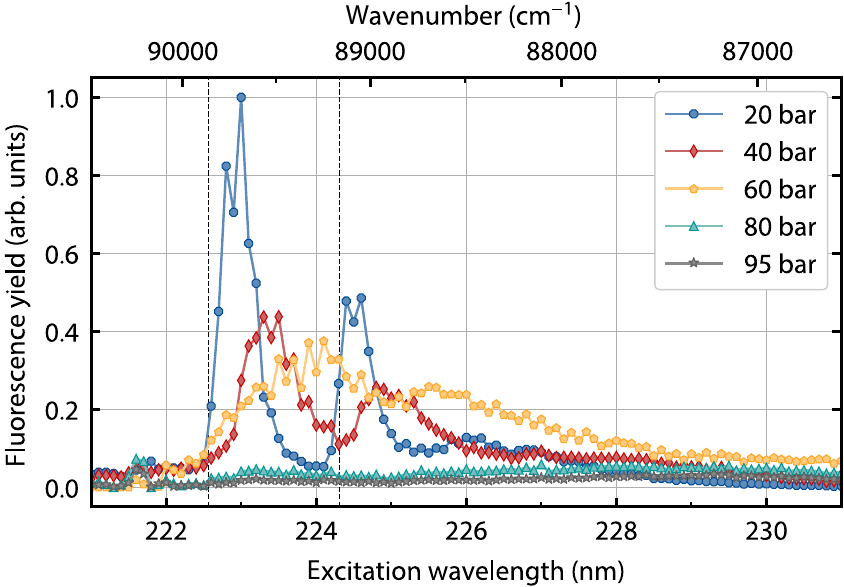}%
		\caption{Excitation spectra of the higher-lying xenon $5p^6 \rightarrow 5p^56p^\prime$ two-photon transitions for different pressures. The vertical dashed lines indicate the expected spectral positions of the undisturbed $5p^6 \rightarrow 5p^56p^\prime[1/2]_0$ and $5p^56p^\prime[3/2]_2$ transitions.
			\label{fig: 6p_prime_resonances}}
	\end{figure}

	Figure \ref{fig: 6p_prime_resonances} illustrates experimental spectra of two-photon transitions to the higher-lying $5p^56p^\prime$ configuration for different xenon pressures, covering the wavelength range from 221 to \SI{231}{\nm}. The vertical dashed lines show the expected $5p^6 \rightarrow 5p^56p^\prime[1/2]_0$ and $5p^56p^\prime[3/2]_2$ transitions at wave numbers  \SI{89860.0}{} and \SI{89162.4}{\per\centi\metre} (\SI{222.6}{} and \SI{224.3}{\nm}) \cite{NISTenergyLevels2023}. The two resonances are clearly observed experimentally, and again a pressure shift as well as a broadening are visible, becoming much more pronounced in the supercritical regime. In general, the pressure broadening of the resonances is less pronounced here than in the case of the $5p^6 \rightarrow 5p^56p$ transitions, enabling the visualization of the data for all recorded pressures in the same plot.
	
\section{Probing Stokes shift compensation with an auxiliary light field}
\label{sec: AbsorptionSpectroscopy}

	In this section, we report results of a scheme devised to compensate for the small reabsorption probability due to the Stokes shift of the emission following one-photon excitation on the one-photon $5p^6 \rightarrow 5p^56s$ transition of xenon near \SI{147}{\nm}. The strong binding of the excited dimer state in dense ensembles leads to the formation of a Xe$_2$ excimer molecule, followed by the spontaneous emission of a highly Stokes-shifted photon. As mentioned above, our aim here is to obtain a large spectral overlap between absorption and emission in this excimer system at room temperature, facilitating the efficient thermalization of vacuum-ultraviolet photons by repeated absorption and emission processes in future microcavity-based experiments.
	
	Here we investigate whether reabsorption of photons around \SI{172}{\nm} can be enhanced by irradiation with an auxiliary light field around \SI{500}{\nm} wavelength, such that excitation into higher levels becomes possible, here specifically in a nondegenerate two-photon transition to the $5p^56p$ manifold [see Fig. \ref{fig: potentialCurves}(c)]. For these investigations, the absorption setup shown in Fig. \ref{fig: experimentalEnvironment}(b) is used. 
	
\begin{figure}
		\includegraphics[]{\figFolder/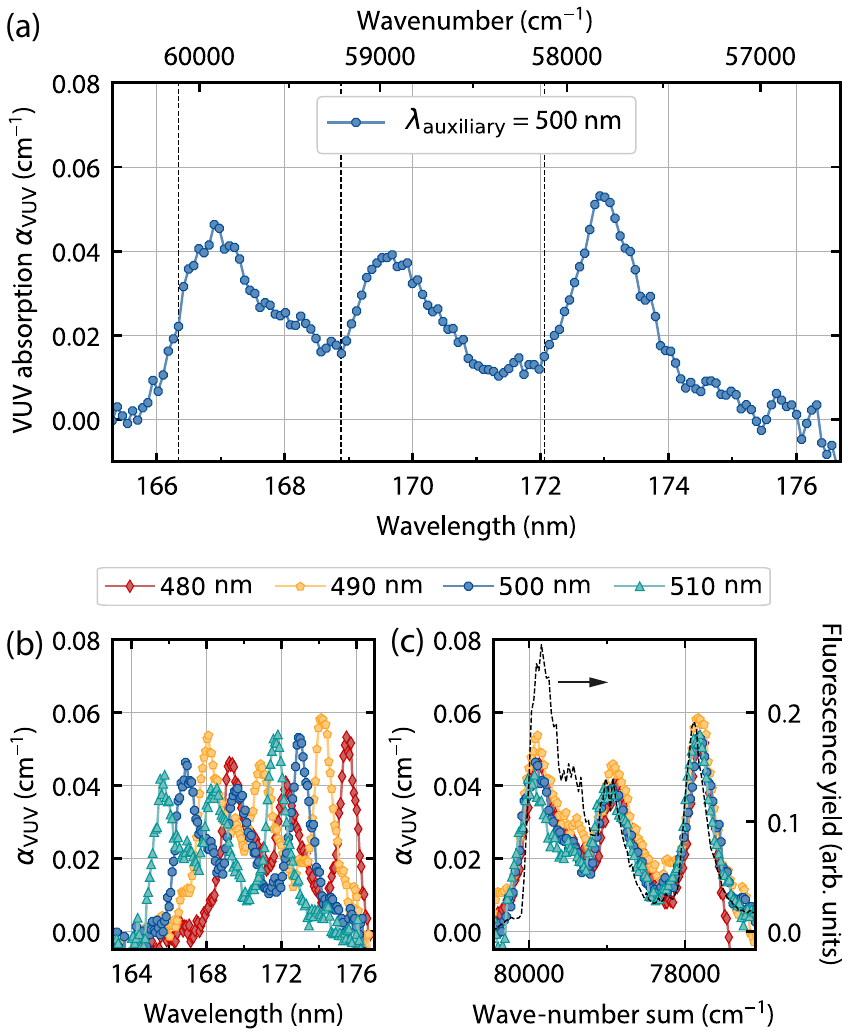}%
		\caption{Absorption spectra of the $5p^6 \rightarrow 5p^56p$ two-photon transitions in xenon for a gas pressure of \SI{35}{\bar}, driven nondegenerately with VUV light in the spectral range near the second excimer continuum and an auxiliary light field. (a) Experimental data, recorded with the auxiliary light field tuned to a wavelength of \SI{500}{\nm}. The plot illustrates the observed variation of the VUV absorption in the gas cell vs wavelength, showing an absorption spectrum of the $5p^6 \rightarrow 5p^56p[1/2]_0$, $5p^56p[3/2]_2$, and $5p^56p[5/2]_2$ two-photon transitions (from left to right). (b) Corresponding data for different wavelengths of the auxiliary light field, resulting in spectrally shifted spectra. (c) The same absorption data as in (b), visualized as a function of the wave-number sum of the two driving light fields, causing the absorption spectra to lie on top of each other. For comparison, the dashed line represents a (degenerate) two-photon excitation spectrum, recorded at the comparable pressure of \SI{40}{\bar} (see, also, Fig. \ref{fig: 6p_resonances}). 
			\label{fig: allMeasurementsTPA}}
	\end{figure}

	Figure \ref{fig: allMeasurementsTPA}(a) shows experimental data for the increase of the absorption of VUV radiation around \SI{172}{\nm} wavelength at a xenon pressure of \SI{35}{\bar}, as induced by irradiation of the auxiliary light field. For this measurement, the gate width of the ICCD camera was set to \SI{8}{\ns}, equaling the pulse length of the OPO source. The used pulse energy was \SI{15}{\mJ}, which in conjunction with a beam diameter of \SI{5}{\mm} (matching that of the overlapping VUV beam) and the pulse length corresponds to an intensity of \SI{96}{\watt\per\square\meter} (average over one pulse). The three visible peaks at around \SI{167}{}, \SI{170}{}, and \SI{173}{\nm} wavelength, for which an absorption signal of the VUV radiation is observed upon irradiation with the auxiliary light field tuned to a wavelength of \SI{500}{\nm}, are attributed to the two-photon transitions $5p^6 \rightarrow 5p^56p[1/2]_0$, $5p^56p[3/2]_2$, and $5p^56p[5/2]_2$, respectively (see, also, our earlier data of Fig. \ref{fig: 6p_resonances} for \textit{degenerate} two-photon spectroscopy with the same upper-state configuration) \cite{NISTenergyLevels2023}. The maximum value of the absorption coefficient of \SI{0.05}{\per\cm} corresponds to an observed reduction of \SI{10}{\percent} of the VUV signal at the corresponding wavelength after passing through the \SI{2}{\cm}-long cell.
		
	To obtain further evidence that the observed VUV absorption spectra are indeed the consequence of a two-photon process, we have conducted measurements at different wavelengths of the auxiliary light field, resulting in spectrally shifted but otherwise unaltered absorption spectra [Fig. \ref{fig: allMeasurementsTPA}(b)]. Figure \ref{fig: allMeasurementsTPA}(c) shows the same spectra visualized as a function of the wave-number sum of the two light fields driving the two-photon transition. As expected, in this visualization, the four signals well lie on top of each other. For comparison, the dashed line represents the corresponding {degenerate} two-photon excitation spectrum for the comparable gas pressure of \SI{40}{\bar} (as shown in Fig. \ref{fig: 6p_resonances}), with the data sets being in good agreement regarding their general shape. It is visible that the relative heights of the three maxima are roughly comparable, while in the case of degenerate two-photon excitation, the $5p^6 \rightarrow 5p^56p[1/2]_0$ transition is more dominant than the other lines. We ascribe this to the different excitation schemes used to drive the respective transitions. The observed slightly larger linewidth in the absorption data is attributed to the inferior wavelength resolution of the gated VUV spectrometer used for these measurements. Our results suggest that the reabsorption probability for VUV photons in the wavelength range around the second excimer continuum can indeed be increased by involving higher-energetic states of the xenon system.
	
\section{Conclusions}
\label{sec: conclusions}

	To conclude, we have reported work investigating two-photon spectroscopy of xenon in the gaseous and supercritical regimes. Both excitation and absorption spectroscopic measurements of the short-wavelength $5p^6 \rightarrow 5p^56p$ and $5p^56p^\prime$ two-photon transitions at pressures of up to \SI{95}{\bar} have been conducted.
	
	Two-photon excitation appears to be a promising avenue to pump future microcavity-based Bose-Einstein condensates of vacuum-ultraviolet photons, as such a scheme allows the pump radiation to be at longer wavelengths than the emission. Typically, for the wavelength range above \SI{200}{\nm}, powerful frequency-converted continuous-wave laser sources are widely available and, e.g., the emission of a frequency-quadrupled Nd laser near \SI{266}{\nm} wavelength may be utilized for pumping via two-photon transitions to the $5p^56p$ excited-state configuration of supercritical xenon.
	
	The key challenge to realize a photon condensate in the vacuum-ultraviolet spectral regime is to identify a suitable thermalization medium. The nondegenerate two-photon absorption measurements performed in this work have illustrated that the large Stokes shift present in the xenon excimer system can be compensated by irradiation of an auxiliary light field. To enhance the magnitude of the reabsorption to a level suitable for the thermalization of photons in a microcavity, an increase of the laser pulse intensity by three to four orders of magnitude would be required, as should be feasible with a stronger focusing of the auxiliary light field to around \SI{100}{\um} beam diameter, which corresponds to the typical size of the thermal photon cloud in present microcavity-based photon BEC experiments. 
	
	A further attractive candidate system for the thermalization of vacuum-ultraviolet photons is constituted by noble gas mixtures \cite{NowakHeteronuclear1985, EfthimiopoulosExcimer1997}, e.g., xenon along with a lighter noble gas in the gaseous or supercritical phase, which, in principle, is not restricted to pulsed operation. In such heteronuclear systems, the excited quasimolecular states are much less strongly bound, resulting in reduced Stokes shifts between absorption and emission spectral profiles. Other candidate systems are mixtures of small heteronuclear molecules, such as the carbon monoxide system, with a noble buffer gas. 

\begin{acknowledgments}
	We acknowledge support of the Deutsche Forschungsgemeinschaft within project WE 1748-25 (Grant No. 684913) and SFB/TR 185 (Grant No. 277625399).
\end{acknowledgments}


%

\end{document}